\documentclass[traditabstract,twocolumn]{aa}
\usepackage{graphicx}
\usepackage{txfonts}
\usepackage{fixltx2e}
\newcommand{\kms}{$\mathrm{km\,s^{-1}}$}
\newcommand{\Egg}{IRAS\,17163}
\newcommand{\Msun}{M$_\odot$}

\begin{document}

\title{ALMA Compact Array observations of the Fried Egg nebula:}
\subtitle{Evidence for large-scale asymmetric mass-loss from the yellow hypergiant \Egg-3907}

\author{S. H. J. Wallstr\"om\inst{1} 
\and E. Lagadec\inst{2}
\and S. Muller\inst{1}
\and J. H. Black\inst{1}
\and N. L. J. Cox\inst{3,4}
\and R. Galv\'an-Madrid\inst{5}
\and K. Justtanont\inst{1}
\and S. Longmore\inst{6}
\and H. Olofsson\inst{1}
\and R. D. Oudmaijer\inst{7}
\and G. Quintana-Lacaci\inst{8}
\and R. Szczerba\inst{9}
\and W. Vlemmings\inst{1}
\and H. van Winckel\inst{10}
\and A. Zijlstra\inst{11}
}

\institute{Department of Earth and Space Sciences, Chalmers University of Technology, Onsala Space Observatory, 439-92 Onsala, Sweden
\\
\email{swallstrom@asiaa.sinica.edu.tw}
\and Laboratoire Lagrange, UMR7293, Université Côte d'Azur, CNRS, Observatoire de la Côte d'Azur, Boulevard de l'Observatoire, 06304 Nice Cedex 4, France 
\and Universit\'{e} de Toulouse, UPS-OMP, IRAP, 31028, Toulouse, France 
\and CNRS, IRAP, 9 Av. colonel Roche, BP 44346, F-31028 Toulouse, France 
\and Instituto de Radioastronom\'ia y Astrof\'isica, UNAM, A.P. 3-72, Xangari, Morelia, 58089, Mexico 
\and Astrophysics Research Institute, Liverpool John Moores University, Liverpool L3 5RF, UK 
\and School of Physics and Astronomy, The University of Leeds, Leeds, LS2 9JT, UK 
\and Instituto de Ciencia de Materiales de Madrid, CSIC, C/Sor Juana In\'es de la Cruz 3, E-28049 Cantoblanco, Spain 
\and Nicolaus Copernicus Astronomical Center, Rabianska 8, 87-100 Torun, Poland 
\and Instituut voor Sterrenkunde, KU Leuven, Celestijnenlaan 200D, 3001 Heverlee, Belgium 
\and Jodrell Bank Centre for Astrophysics, Alan Turing Building, Manchester M13 9PL, UK 
}

\date{Received / Accepted}
   
\titlerunning{ALMA ACA observations of the Fried Egg}
\authorrunning{S. H. J. Wallstr\"om et al.}
 
\abstract
{
Yellow hypergiants are rare and represent a fast evolutionary stage of massive evolved stars. That evolutionary phase is characterised by a very intense mass loss, the understanding of which is still very limited. Here we report ALMA Compact Array observations of a 50$''$-mosaic toward the Fried Egg nebula, around one of the few Galactic yellow hypergiants \Egg-3907. The emission from the $^{12}$CO J=2-1 line, H30$\alpha$ recombination line, and continuum is imaged at a resolution of $\sim$8$''$, revealing the morphology of the molecular environment around the star. The continuum emission is unresolved and peaks at the position of the star. The radio recombination line H30$\alpha$ shows unresolved emission at the star, with an approximately gaussian spectrum centered on a velocity of 21$\pm$3~\kms\ with a width of 57$\pm$6~\kms. In contrast, the CO 2-1 emission is complex and decomposes into several components beyond the contamination from interstellar gas in the line of sight. The CO spectrum toward the star is a broad plateau, centered at the systemic velocity of +18 \kms\ and with an expansion velocity of 100$\pm$10 \kms. Assuming isotropic and constant mass-loss, we estimate a mass-loss rate of 8$\pm$1.5 $\times10^{-5}$~M$_\odot$\,yr$^{-1}$. At a radius of 25$''$ from the star, we detect CO emission associated with the dust ring previously imaged by {\it Herschel}. The kinematics of this ring, however, is not consistent with an expanding shell, but show a velocity gradient of $v_{sys} \pm$20 \kms. In addition, we find a puzzling bright feature radially connecting the star to the CO ring, at a velocity of +40 \kms\ relative to the star. This spur feature may trace a unidirectional ejection event from the star. Our ACA observations reveal the complex morphology around \Egg\ and illustrate the breakthroughs that ALMA will bring to the field of massive stellar evolution.
}


\maketitle
%

\section{Introduction}

Massive stars, with initial masses between 8 and 40~\Msun, spend only $\sim$10 million years on the main sequence before becoming large and cool red supergiant (RSG) stars. The most massive of these stars, M$>$20~\Msun, will then evolve via a short-lived (10$^{2-3}$ years) yellow hypergiant (YHG) phase, followed by a luminous blue variable phase, to finally become Wolf-Rayet stars \citep{Oudmaijer:2009aa}.
The intense mass-loss associated with these phases, and the eventual supernova explosion, provides kinetic energy and chemical enrichment to the surrounding ISM. The stellar winds lead to the formation of a circumstellar envelope (CSE), that will affect the shape and evolution of the future supernova remnant, and determine how the material will be incorporated into the ISM. The CSE also traces the mass-loss history of the star, providing valuable information about the past evolution and possible future of these rare giants.

\Egg-3907 (also called Hen 3-1379 or the Fried Egg nebula) is one of the brightest mid-infrared sources in the sky and was initially classified as a post-AGB star by \citet{Szczerba:2007aa}. More recently, the distance (4$\pm$0.5 kpc) to this object has been well constrained by the presence of interstellar absorption lines in its optical spectra \citep{Lagadec:2011aa}, implying that the star is four times further away than previously assumed. It is thus too luminous to be a post-AGB star. The location of IRAS 17163 on a temperature-luminosity diagram and the similarity of its optical spectrum with the well-studied YHG IRC+10420 \citep{Wallstrom:2015aa} suggest that its properties are very similar to the brightest YHGs \citep{de-Jager:1998aa}. Hence, IRAS 17163 is one of the rare Galactic yellow hypergiants. These stars are known to have circumstellar envelopes of gas and dust, but only two other objects, IRC+10420 and AFGL~2343, have an extended molecular envelope \citep{Castro-Carrizo:2007aa}. Both objects show an overall spherical symmetry, and evidence of a detached shell.

Resolved images of \Egg\ and its envelope \citep{Lagadec:2011aa}, obtained using the VLT mid-infrared instrument VISIR, allowed the dust distribution around the central star to be mapped. Two concentric spherical dusty shells were resolved within 2.5$''$ of the star (note that at a distance of 4~kpc, 1$''$ corresponds to $\sim$0.02 pc or 4000 AU). Radiative transfer modeling of the circumstellar environment indicates the presence of 0.04~M$_{\odot}$ of dust in the envelope, ejected during the last few hundred years. Furthermore, a larger, cooler dust shell with a radius $\sim$25$''$ was discovered with the {\it Herschel Space Observatory} \citep{Hutsemekers:2013aa} and they estimate the total mass of gas and dust, including this shell, to be $\sim$7~M$_{\odot}$ (assuming a gas-to-dust mass ratio of 40).

To further constrain the properties of the circumstellar envelope of \Egg, particularly the mass-loss rate and kinematics, \citet{Wallstrom:2015aa} performed spectral observations of the CO J=2-1 and 3-2 lines with the Atacama Pathfinder Experiment (APEX) telescope. They find a complex CO spectrum with asymmetric line profiles and multiple velocity components in addition to the contamination by interstellar gas in the line of sight. However, the limited spatial resolution and sensitivity prevented them from fully interpreting the complex emission. 

We have obtained continuum and CO 2-1 Atacama Large Millimeter/Submillimeter Array (ALMA) compact array (ACA) observations that allow us to shed light on the morphology and kinematics of the circumstellar environment of \Egg. These observations are presented in Section 2, followed by Results in Section 3, a Discussion in Section 4, and finally Conclusions in Section 5.

\section{Observations and data reduction}

\subsection{ALMA Compact Array data}

\Egg\ was observed as part of project 2013.1.00502.S with the ALMA Compact Array (ACA) between the 25$^\mathrm{th}$ and 27$^\mathrm{th}$ of July 2014, in a mosaic of seven pointings to cover a field of $\sim$50\arcsec. The mosaic was centered on R.A. 17:19:49.335 and Dec $-$39:10:37.94 (J2000), with the six additional pointings offset by $\sim$28$''$ in a hexagonal pattern. 
The array configuration and $uv$-coverage resulted in a synthesized beam of $8.4'' \times 4.3''$ (with a position angle of 96.5$^\circ$) at the frequency of the CO 2-1 line, using natural weighting. 
The projected baselines ranged between $\sim$5 and 45~m, giving a maximum recoverable scale of $\sim$30$''$.  

Two spectral windows 1.875~GHz wide were set up with 0.977~MHz channel spacing, centered at 230.481~GHz (upper side band, covering the CO 2-1 transition) and 217.047~GHz (lower side band, covering the SiO 5-4 transition, which was not detected). The velocity resolution for these windows is 1.7~\kms\ after Hanning smoothing. In addition, two more 2-GHz windows were set up with a coarse spectral resolution of 15.6 MHz for continnum imaging, centered at frequencies of 215.9 and 231.5~GHz. We detect emission from the H30$\alpha$ line in the latter. 

Data were calibrated and imaged in a standard way using CASA\footnote{http://casa.nrao.edu/}. The Jy-to-K conversion factor to brightness temperature for the CO 2-1 data is 1~Jy = 0.637~K. Continuum was subtracted from the line emission data by using the CASA task {\it uvcontsub} and fitting a first order polynomial to line-free channels.

\subsection{APEX data}

Complementary APEX\footnote{This publication is based on data acquired with the Atacama Pathfinder Experiment (APEX). APEX is a collaboration between the Max-Planck-Institut f\"ur Radioastronomie, the European Southern Observatory, and the Onsala Space Observatory.} observations were carried out with the APEX-1 receiver \citep{Vassilev:2008aa}
on August 27th, 2015 under very good weather conditions (precipitable water vapor of $\sim$0.8~mm). The $^{13}$CO 2-1 transition at 220.4~GHz was observed in a single pointing centered on IRAS~17163 (main beam 28"), with the same fixed off-position used by \citet{Wallstrom:2015aa}, $\sim$40$''$ south-west of the star. The data were reduced in the standard single-dish data reduction package CLASS\footnote{http://www.iram.fr/IRAMFR/GILDAS}.

\begin{figure}[t!]
  \begin{center}
  \includegraphics[width=\columnwidth]{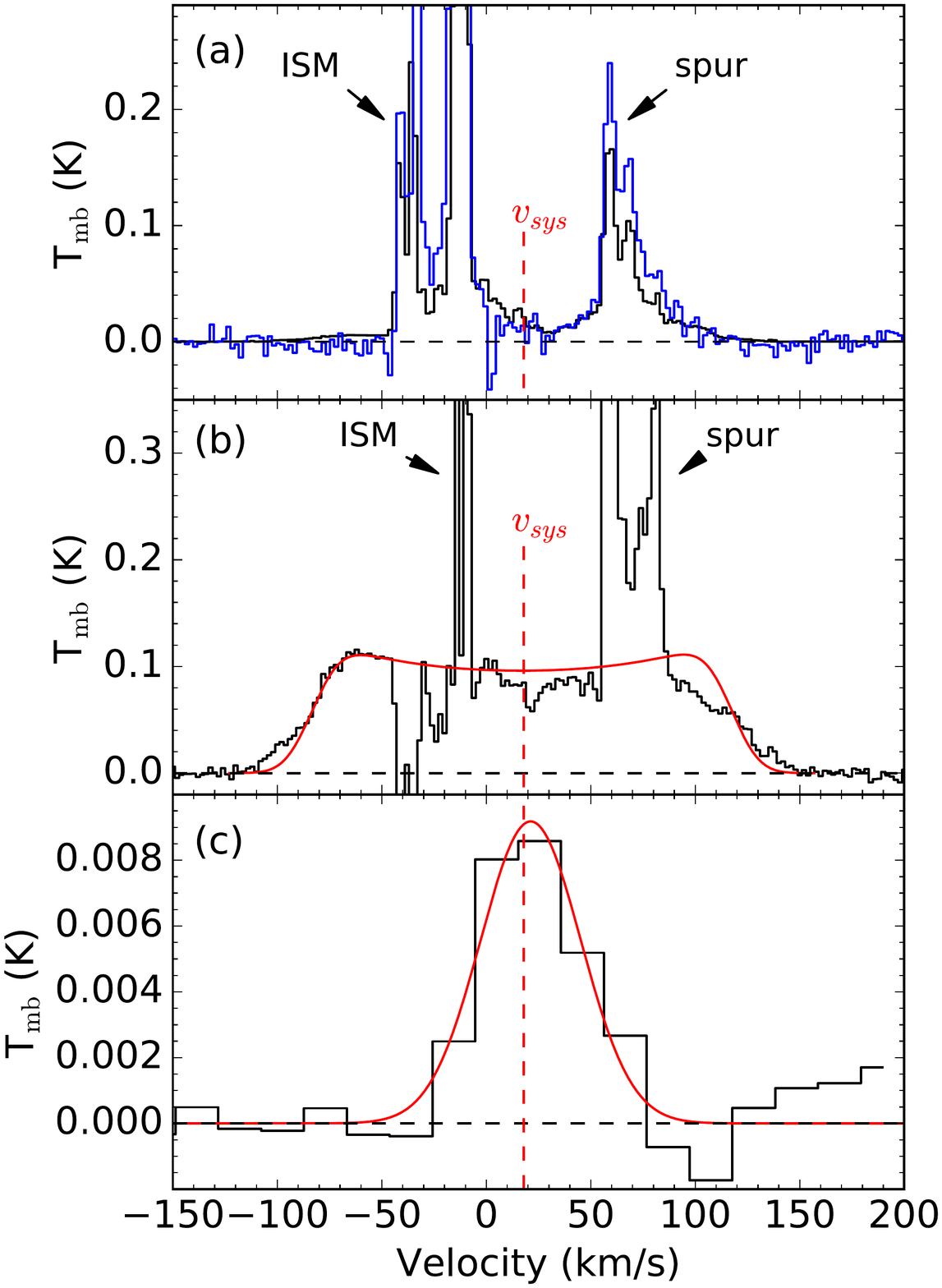}
  \end{center}
 \caption[]{{\it (a)}~In black, a spectrum of the CO 2-1 ACA data smoothed to the 27$''$ beam of APEX. In blue, the APEX CO 2-1 spectrum from \citet{Wallstrom:2015aa}.  {\it (b)}~ACA CO~2-1 spectrum summed over the synthesised beam at the position of IRAS~17163, showing the broad plateau seen only towards the star. The best fit mass-loss model is overlaid in red. {\it (c)}~The~H30$\alpha$ spectrum, averaged over the synthesised beam at the position of \Egg. The best fit gaussian line profile is overlaid in red, centered on 21 \kms\ with a FWHM of 57 \kms.
 
In all spectra the systemic velocity at +18 \kms\ (see Sect.~\ref{sect:wind}) is marked with a dashed red line, and in spectra {\it a} \& {\it b} the absorption-like features between -50 and 0 \kms\ are artefacts caused by the filtering out of extended emission. 
\label{fig:spec3}}
\end{figure}

 \begin{figure*}[ht!]
  \begin{center}
    \includegraphics[width=0.7\textwidth,angle=90]{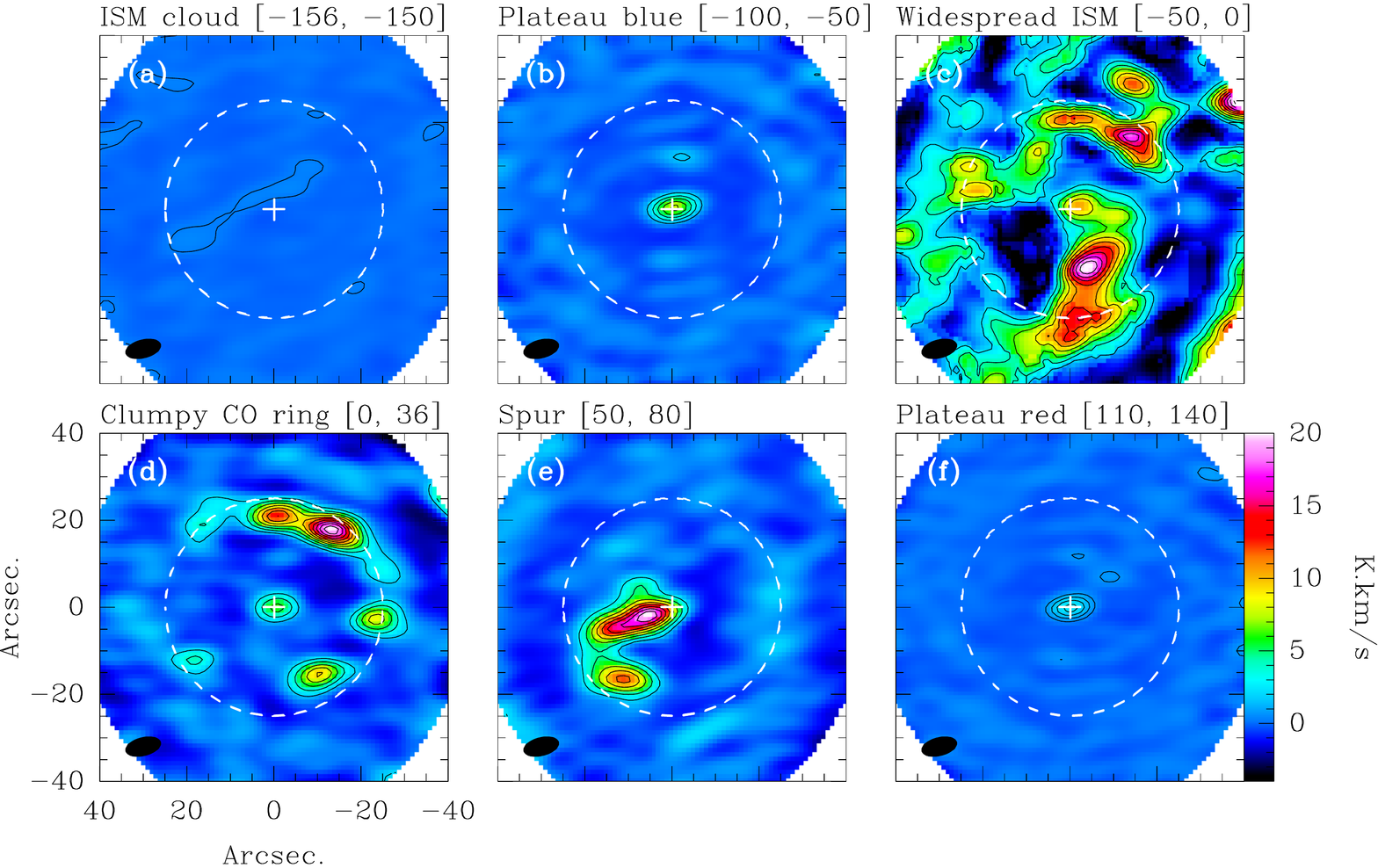}
  \end{center}
 \caption[]{Integrated CO 2-1 maps over several different features in the ACA data, with the velocity range of each map given in brackets. North is up and East is to the left in each map. {\it (a)} An interstellar cloud around $-150$~\kms. Contours start at 4$\sigma$ and are spaced by 4$\sigma$ ($\sigma$=57~mK\,\kms). {\it (b)} The blue edge of the stellar wind plateau between $-100$ and $-50$ \kms. Contours start at 10$\sigma$ and are spaced by 10$\sigma$ ($\sigma$=146~mK\,\kms). {\it (c)} Widespread ISM emission between $-50$ and 0~\kms. Contours start at 15$\sigma$ and are spaced by 15$\sigma$ ($\sigma$=140~mK\,\kms). {\it (d)} The stellar wind at the star, and the clumpy CO ring, integrated between 0 and +36~\kms, symmetrically around the systemic velocity. Contours start at 15$\sigma$ and are spaced by 15$\sigma$ ($\sigma$=124~mK\,\kms). {\it (e)} The spur, integrated between 50 and 80~\kms. Contours start at 20$\sigma$ and are spaced by 20$\sigma$ ($\sigma$=114~mK\,\kms). {\it (f)} The red edge of the stellar wind plateau between +110 and +140~\kms. Contours start at 5$\sigma$ and are spaced by 5$\sigma$ ($\sigma$=114~mK\,\kms).

The restoring beam (8.4$''\times$4.3$''$) is shown as a filled ellipse in the bottom left of each map, and the common color scale is shown in the bottom right map. The location of the star is marked with a cross and the dust shell seen with {\it Herschel} is represented by a dashed circle with 25$''$ radius.
\label{fig:mom0}}
\end{figure*}

\section{Results}

In this section, we describe the morphology of the CO 2-1 emission revealed by the ACA. We identify several distinct features in the data, which is shown smoothed to the resolution of APEX in Fig.~\ref{fig:spec3}{\it a}. Between $-50$ and 0 \kms\ we see widespread ISM emission, around the systemic velocity at +18 \kms\ (see Sect.~\ref{sect:wind}) is emission from a clumpy CO ring, between +50 and +80 \kms\ is a bright feature we call the "spur", and between 90 and 110 \kms\ is emission from a western arc. Each of these components is discussed in a separate section below. 

In addition, we find the H30$\alpha$ recombination line of hydrogen at 231.9~GHz as a spatially unresolved feature centered on \Egg. The line profile is resolved with a velocity resolution of $\sim$20~\kms, and a gaussian fit finds the line centered on 21$\pm$3~\kms\ with a FWHM of 57$\pm$6~\kms\ and a peak brightness temperature of $\sim$9~mK  (Fig.~\ref{fig:spec3}{\it c}).

Fig.~\ref{fig:mom0} presents integrated maps over various features. The dashed circle represents the approximate location of the "{\it Herschel} ring", a cool dust ring seen with {\it Herschel} \citep{Hutsemekers:2013aa}, with a radius of 25$''$ and centered on IRAS\,17163. All velocities are given in the local standard restframe (LSR). Channel maps between $-$100 and +150~\kms\ are also presented in Fig.~\ref{fig:chanmap}.

 \begin{figure}
  \begin{center}
  \includegraphics[width=\columnwidth]{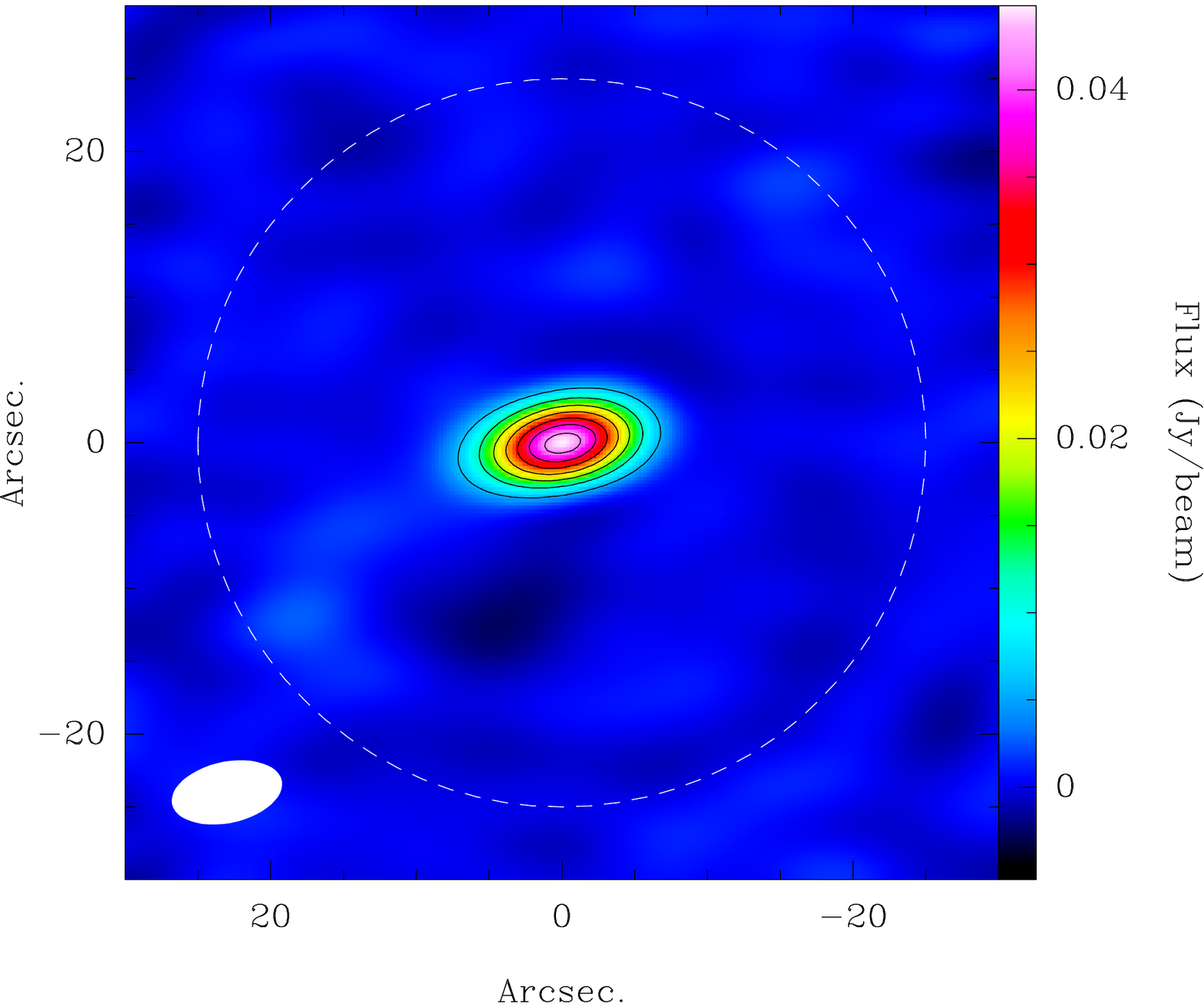}
  \end{center}
 \caption[]{The ACA continuum emission, showing the central unresolved feature. Contours start at 10$\sigma$ ($\sigma$=0.6 mJy/beam) and are spaced by 10$\sigma$. The location of the {\it Herschel} dust shell is shown with a dashed 25$''$-radius circle, and the restoring beam is shown as a white ellipse in the bottom left. 
\label{fig:cont}}
\end{figure}

\subsection{Continuum emission}
The continuum emission (Fig.~\ref{fig:cont}), obtained by aggregating all line-free channels in the coarse spectral resolution windows, is an unresolved feature at the location of the star, with flux densities of 42 and 48 mJy in the two bands centered on 215.9 and 231.5~GHz, respectively. 
SED fitting by \citet{Hutsemekers:2013aa} found two dust components at temperatures of $\sim$180 and 60~K. However, their longest wavelength observations are 160~$\mu$m ($\simeq$1880 GHz), so the cold dust is not well constrained. Extrapolating their SED fit (modified blackbody emission with $\beta$=2) gives fluxes $<$20 mJy around 230 GHz. We find values a factor of $\sim$2 larger, suggesting there may be a colder dust component within 8$''$ of the star.

\subsection{The interstellar surroundings}

\Egg\ is in a complex region of the sky, with widespread CO \citep{Arnal:2008aa} and cool dust \citep{Hutsemekers:2013aa} emission. 
Indeed, the brightest CO emission in the ACA dataset is ISM emission which covers the field of view between roughly $-$50 and 0 \kms, as seen in Figs.~\ref{fig:spec3}{\it a} \& \ref{fig:mom0}{\it c}. The contamination from these ISM features precludes the analysis of any potential circumstellar emission at velocities in this range. 
There are two distinct ISM components, centered around $-38$ and $-12$ \kms, which match the CO 1-0 emission seen towards the nearby star forming regions RCW 121 and RCW 122 \citep{Arnal:2008aa}. These velocities are consistent with Galactic rotation in the direction of IRAS~17163 at distances of 4 and 1~kpc, respectively \citep{Reid:2009aa}.
In addition, we note a fainter diffuse filament around $-150$~\kms\ (see Fig.~\ref{fig:mom0}{\it a}), which is consistent with Galactic rotation at 7~kpc. 

We note that our mosaic does not cover the full extent of the widespread ISM emission around $-38$ and $-12$ \kms. This causes negative artefacts in the data, due to the filtering out of extended emission ($>30''$) by the interferometric observations.

\subsection{The stellar wind}\label{sect:wind}

Figure~\ref{fig:spec3}{\it b} shows the CO 2-1 spectrum extracted toward \Egg. It was obtained by summing the emission within the synthesized beam centered on the star.
In this spectrum, the most prominent emission feature (v$\sim$$-$10~\kms) corresponds to interstellar CO, as discussed above. Next is the bright feature between $\sim$50 and 80~\kms\ that was observed in previous APEX spectra \citep{Wallstrom:2015aa}. 
We discuss this feature, which we call the spur, hereafter in Sect.~\ref{sect:spur}.
Finally, clearly detected with the ACA, is a faint ($\sim$0.1~K in the ACA beam) plateau centered on a velocity of $\sim$20~\kms\ and with a broad width of $\sim$250~\kms, seen only towards the star (see Fig.~\ref{fig:mom0}{\it b,d,f}). This resolves the previously found velocity discrepancy between APEX CO peak emission around +60 \kms\ and optical Fe\,II emission lines at +18 \kms\ \citep{Wallstrom:2015aa}. Henceforth we take the systemic velocity of \Egg\ to be +18 \kms.

Assuming the plateau corresponds to a stellar wind, we use a simple model of an expanding circumstellar envelope where the radiative transfer is solved using a Monte Carlo method \citep{Bernes:1979aa,Schoier:2001ab} to constrain the mass-loss rate. This model assumes that the envelope expands at a constant velocity and is formed by a constant, isotropic, and smooth mass-loss. It incorporates CO rotational energy levels up to J=40 for the vibrational ground state and the first vibrationally excited state, and uses collisional cross-sections from \citet{Yang:2010aa}. 
In our modeling we take the stellar parameters (distance, effective temperature, luminosity) from \citet{Lagadec:2011aa}, 
and assume a [CO]/[H$_2$] ratio of 10$^{-4}$, a turbulent velocity of 1~\kms, and a gas-to-dust ratio of 40 (cf. \citealp{Hutsemekers:2013aa}). 
For our parameters the model finds a CO half-abundance radius (determined by photodissociation) of 1.55$\times$10$^{17}$~cm, corresponding to 2.6$''$ at 4~kpc, consistent with the extent of the dust shells found by \citet{Lagadec:2011aa}.

In order to fit the mass-loss rate and expansion velocity, we executed a grid of models over the parameter space and determined the best fit model by  $\chi^2$ minimisation to the parts of the spectrum not contaminated with ISM or spur emission. We find an expansion velocity of 100$\pm$10~\kms\ and a mass-loss rate of 8$\pm$1.5 $\times10^{-5}$~M$_\odot$\,yr$^{-1}$ (Fig.~\ref{fig:spec3}{\it b}). These errors are the statistical errors corresponding to the 1$\sigma$ $\chi^2$ error ellipse. 
The mass-loss rate estimate has further uncertainty associated with the model parameters, such as the distance to the star and the assumption of constant and isotropic mass-loss, which are not taken into account here. Integrating over the model fit we find a total integrated intensity of $\sim$21~K\,\kms\ for the wind.

We note that the model was developed for the circumstellar envelopes of AGB stars, and hence may not be strictly applicable to our YHG star. For example, it assumes an entirely molecular gas envelope, whereas the envelope of \Egg\ might be significantly atomic due to the high stellar temperature. On the other hand, this method provides an order of magnitude estimate of the mass-loss rate, allowing us to make comparisons to other massive evolved stars.
Accordingly, our estimated mass-loss rate is similar to that of the archetypal YHG IRC+10420, of $\sim$$1-9\times10^{-4}$~M$_\odot$\,yr$^{-1}$ \citep{Castro-Carrizo:2007aa,Dinh-V.-Trung:2009aa}. 

From the expansion velocity, CO half-abundance radius, and (assumed constant) mass-loss rate we can estimate a total gas mass for this stellar wind of 0.04~\Msun. We note that this estimate assumes an entirely molecular wind and hence, as there may be additional mass outside the extent of the CO envelope, it should be viewed as a lower bound on the total gas mass.  Our estimate is similar to the mass found in the inner 2.5$''$ region by \citet{Hutsemekers:2013aa} of 0.08~\Msun, assuming a gas-to-dust ratio of 40. On the other hand, \citet{Lagadec:2011aa} find a much larger mass in this region: a total dust mass of 0.04~\Msun, the bulk of which is in the inter-shell region between two concentric dust shells.
Note, however, that the inter-shell dust was added, and the outer radius in the \citet{Lagadec:2011aa} model was arbitrary extended to about 5$''$, in an attempt to explain the IRAS fluxes which we now know are mostly due to the emission from the {\it Herschel} dust shell \citep{Hutsemekers:2013aa}. Hence, updating the \citet{Lagadec:2011aa} modelling of the inner shells, without the IRAS fluxes and with better constraints on the radial boundaries, results in a total dust mass of 0.003~\Msun\ which, assuming a gas-to-dust ratio of 40, corresponds to a gas mass of 0.12~\Msun. Given the uncertainties in the models of both gas and dust, this is in reasonable agreement with our lower bound estimate of 0.04~\Msun.

The expansion velocity of 100$\pm$10~\kms\ is unusually high for a post-RSG star: \citet{Oudmaijer:2009aa} found typical velocities of $\sim$20 \kms. They found only two objects with similar outflow velocities, the warm supergiants HD~101584 and Frosty Leo, both of which have bipolar outflows. Some OH/IR stars with bipolar outflows also show similarly large expansion velocities \citep{Zijlstra:2001aa}. Although it is not known what powers such high velocity outflows, they seem more likely to be associated with a bipolar flow. However, the central emission associated with the fast stellar wind of \Egg\ is unresolved in the ACA beam so its geometry remains elusive.

\subsection{The clumpy CO ring}\label{sect:ring}

\begin{figure} 
  \begin{center}
\includegraphics[width=\columnwidth]{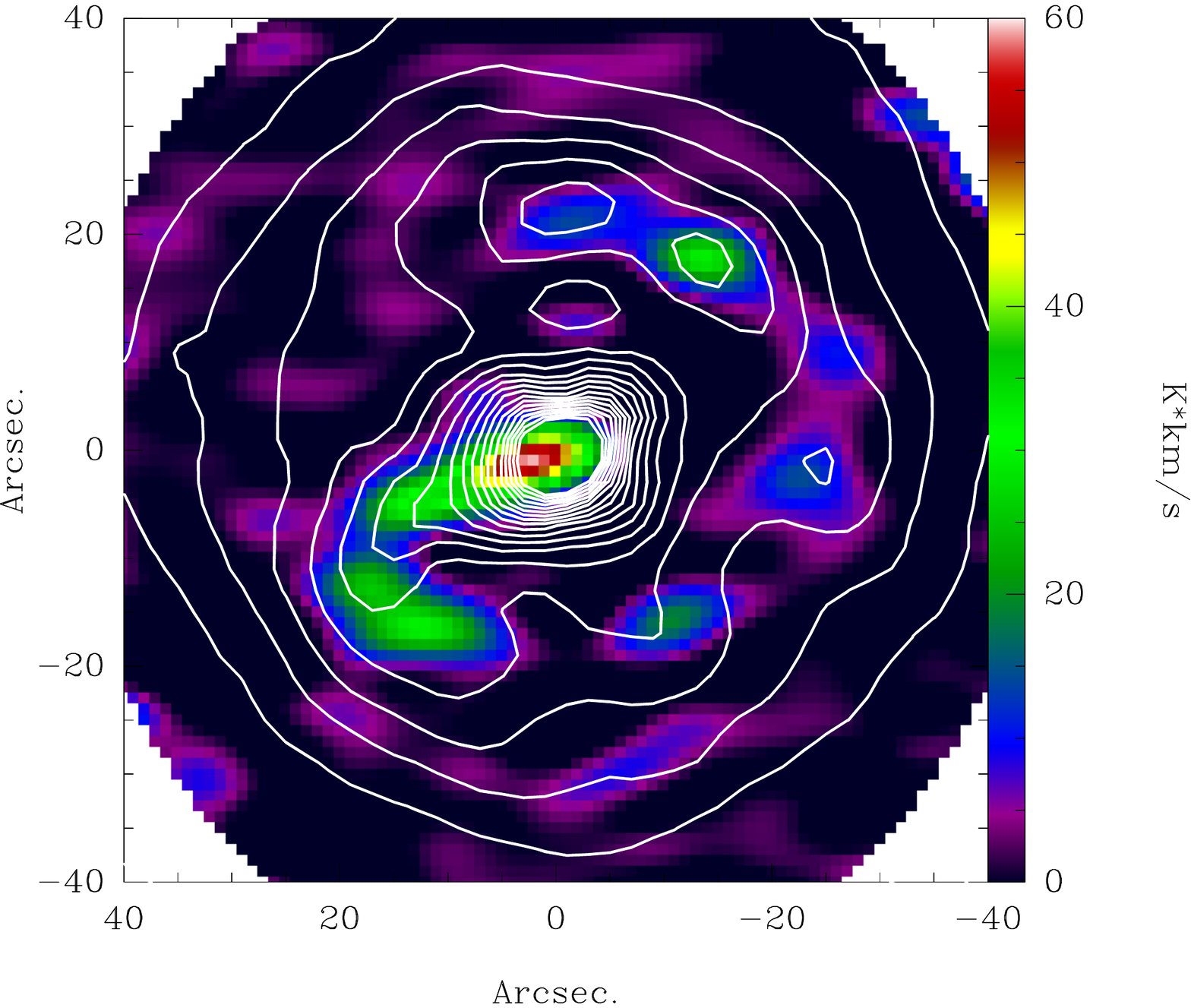}
  \end{center}
 \caption[]{In color: ACA CO 2-1 emission integrated over the velocity range 0 to +140~\kms\ (avoiding the ISM contamination). Note that the emission peaks not at the star but on the spur feature. Overlaid in gray are {\it Herschel} 70 $\mu$m dust contours (see \citealp{Hutsemekers:2013aa}).
 \label{fig:contours}}
\end{figure}

\begin{figure}
  \begin{center}
\includegraphics[width=\columnwidth]{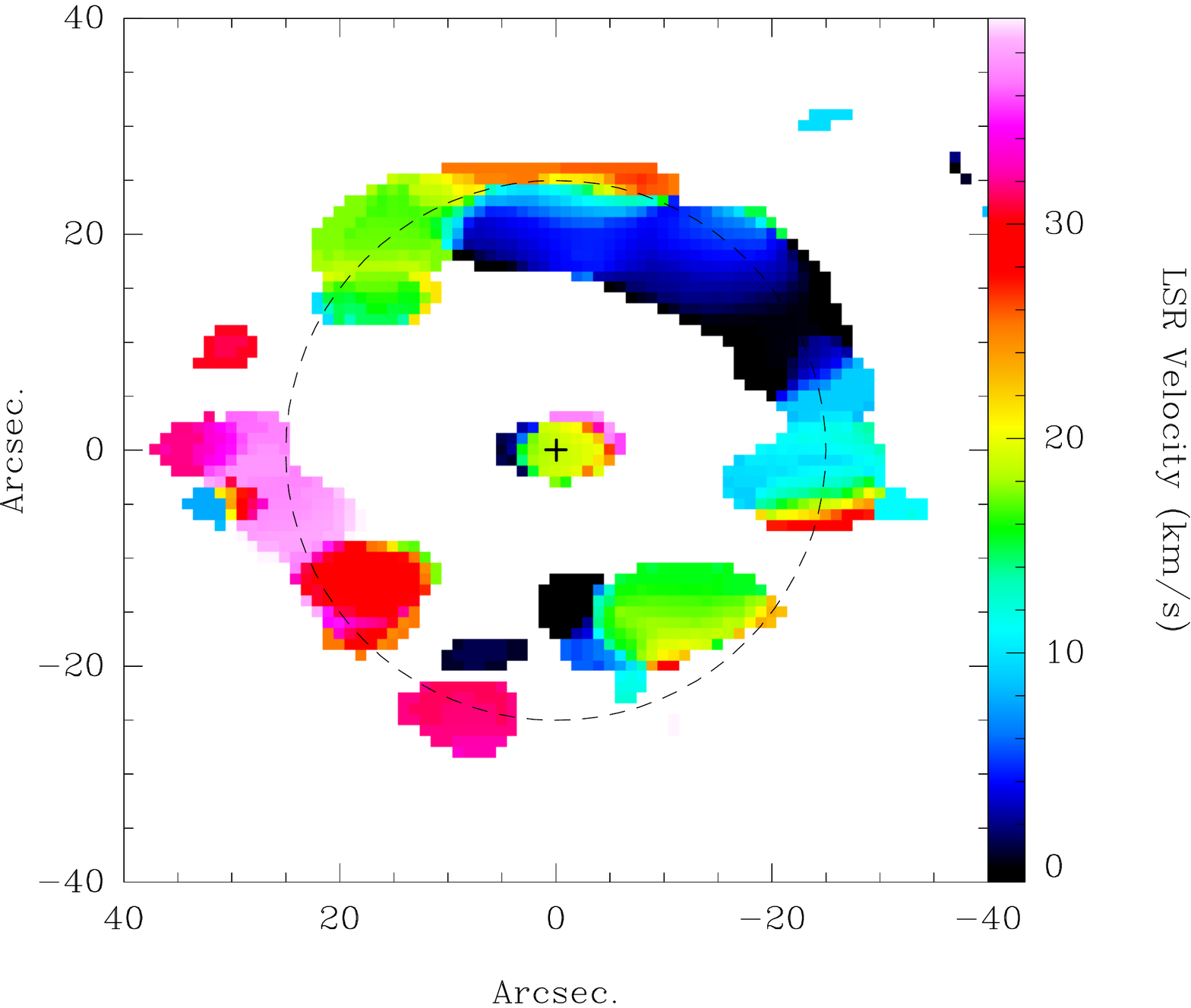}
  \end{center}
 \caption[]{An intensity weighted mean velocity map of the clumpy CO ring integrated within 20 \kms\ of the systemic velocity at $+18$\kms, i.e. between $-2$ and $+38$ \kms, to avoid any confusion from the ISM or spur features. The location of the star is marked with a cross, and the location of the {\it Herschel} dust ring is indicated with a dashed circle of 25$''$ radius.
\label{fig:vel_center}}
\end{figure}

In addition to the central unresolved emission from the stellar wind, the CO channel maps show a series of clumps whose overall distribution reproduces the cool (T$\sim$60~K) dust ring seen with {\it Herschel} (\citealp{Hutsemekers:2013aa}; see Fig.~\ref{fig:contours}). These clumps have a wide range of velocities but the contamination with interstellar CO (at $v<0$~\kms) on one side and the spur ($v \sim 50-80$~\kms) on the other makes it difficult to define the total extent in velocity. In any case, the overall kinematics clearly do not reflect that of an expanding shell. A shell expanding at constant velocity would show up in the channel maps (Fig.~\ref{fig:chanmap}) projected as a ring changing in size with velocity and reaching a maximum extent at the systemic velocity. Instead, the velocity structure of the clumpy CO ring shows a strong velocity gradient with red-shifted clumps to the south-east and blue-shifted clumps to the north-west (Fig.~\ref{fig:vel_center}). Note that Fig.~\ref{fig:vel_center} shows a restricted velocity range of $\pm20$~\kms\ around the systemic velocity to avoid confusion from the ISM or spur features. The integrated intensity of this clumpy ring, between 0 and 36 \kms, is $\sim$65~K\,\kms.

In order to estimate the masses of the CO clumps, we used the radiative transfer code RADEX \citep{van-der-Tak:2007aa} to find the range of parameters needed to reproduce the average peak clump brightness in CO 2-1, of 1.1$\pm$0.7~K. To this end we fixed the linewidth to 10 \kms\ and the background temperature to 2.73~K, and explored a large parameter space over kinetic temperature, H$_2$ density, and CO column density. The temperature and H$_2$ density could not be constrained, but the CO column density was found to be 10$^{16-17}$~cm$^{-2}$, which corresponds to a mass of 0.03-0.3~M$_\odot$ per clump (assuming a spherical clump diameter of 8'' and a CO abundance of 10$^{-4}$ compared with H$_2$). We identify at least 8 separate clumps and hence a minimum total mass in the ring of $\sim$0.2-2~M$_\odot$.

\subsection{The spur}\label{sect:spur}

In the velocity range $\sim$+50 to +80~\kms\ (LSR) there is a bright elongated CO feature we call the spur, extending radially from \Egg\ towards the south-east (position angle of $112^\circ$) to the clumpy CO ring. It is redshifted with respect to the star by $\sim$30--60 \kms. This component corresponds to the emission seen in the APEX spectra by \citet{Wallstrom:2015aa} and reaches a peak brightness temperature of 7.8~K in the ACA beam. The integrated intensity, between 50 and 80 \kms, is $\sim$83~K\,\kms. The spectrum of the spur shows a bright peak at +58 \kms\ and fainter peaks at +66 and +80 \kms. The spatial structure of the spur is complex, seeming to break up into several components between +65 and +80 \kms\ (see Fig.~\ref{fig:chanmap}). We also note that the spur appears in the {\it Herschel} dust contours (Fig.~\ref{fig:contours}), suggesting it contains dust at a temperature of 60~K. To obtain a rough estimate of the mass of the spur we apply the same method as described in Sect.~\ref{sect:ring} to find $N_{CO}$=10$^{18}$~cm$^{-2}$, corresponding to a gas mass of $\sim$3~M$_\odot$ for an ellipse with major and minor axes of $\sim$15$''$ and 5$''$, respectively.

\begin{figure}[t]
  \begin{center}
   \includegraphics[width=\columnwidth]{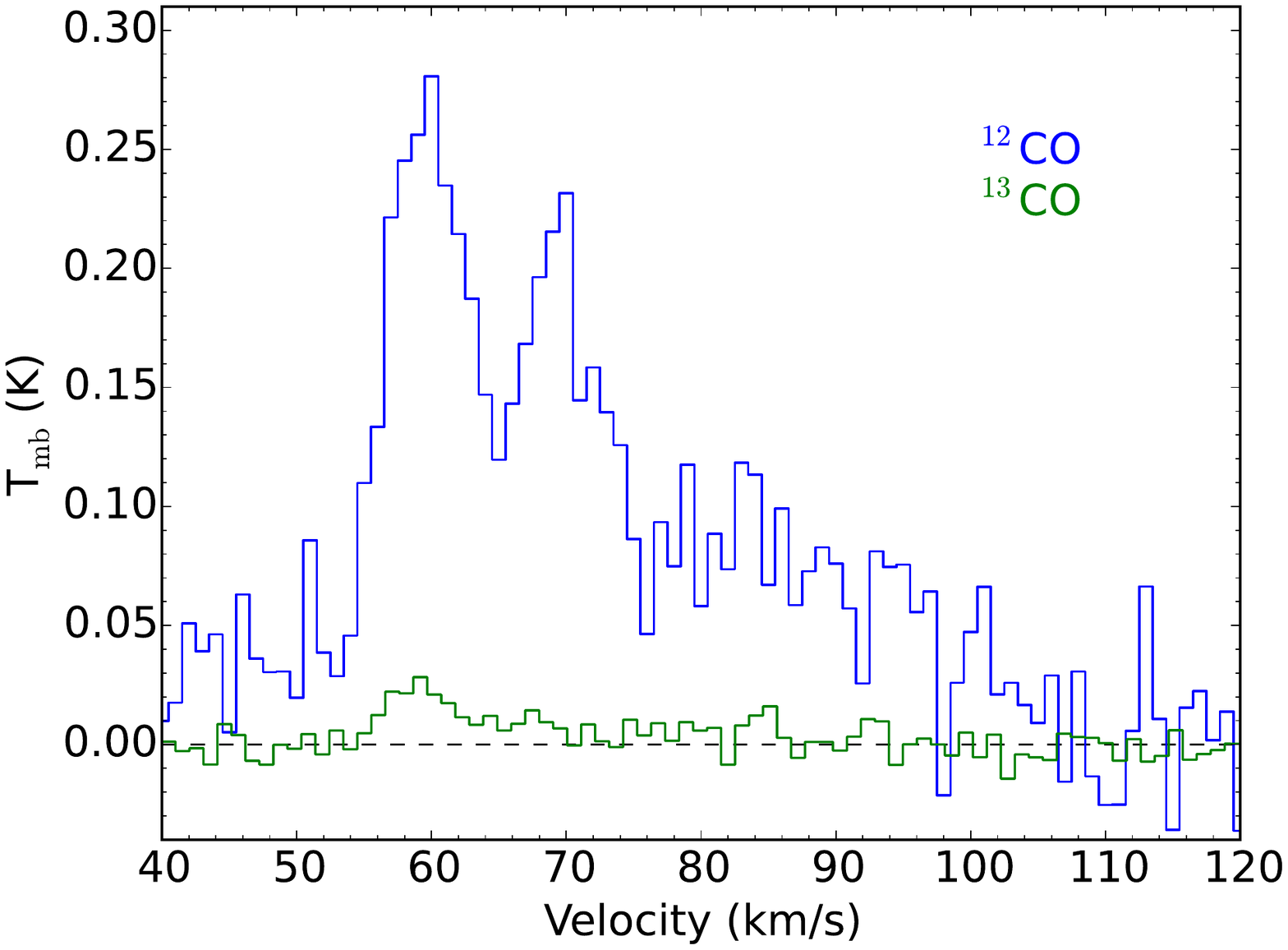}  
  \end{center}
 \caption[]{APEX observations of $^{12}$CO (blue) and $^{13}$CO (green) 2-1 emission around IRAS~17163.
\label{fig:13co}}
\end{figure}

One way to determine if the spur is associated with the star is to measure the $^{12}$C/$^{13}$C ratio. We obtained an APEX spectrum in $^{13}$CO 2-1 (Fig.~\ref{fig:13co}) and compared with the $^{12}$CO spectrum from \citet{Wallstrom:2015aa}. The $^{13}$CO spectrum shows peaks coincident with the $^{12}$CO peaks, but it is very faint with a peak brightness of 0.02~K, corresponding to a signal-to-noise ratio of 4. In order to estimate the $^{12}$CO/$^{13}$CO ratio we fit both spectra simultaneously with the same profile, multiplied by a scaling factor. The profile was well decomposed into three gaussian components.
We find a $^{12}$CO/$^{13}$CO intensity ratio of $13^{+10}_{-5}$. Equating this to the $^{12}$C/$^{13}$C abundance ratio requires the assumption that $^{12}$CO is optically thin. This is likely, given that the spur is resolved in the ACA data and has a peak temperature of $\sim$8 K, far below the 60 K temperature of the associated dust. 
A $^{12}$C/$^{13}$C abundance ratio around 10 is indicative of material ejected from an RSG star \citep{Gonzalez:2000aa,Origlia:2013aa}, while the ISM ratio at the galactic radius of \Egg\ is around 40 (\citealp{Wilson:1994aa}; assuming a distance of 4 kpc).
Hence, this material was most likely ejected by the star in its previous RSG phase.

\subsection{Western arc}

\begin{figure}[t]
  \begin{center}
   \includegraphics[width=\columnwidth]{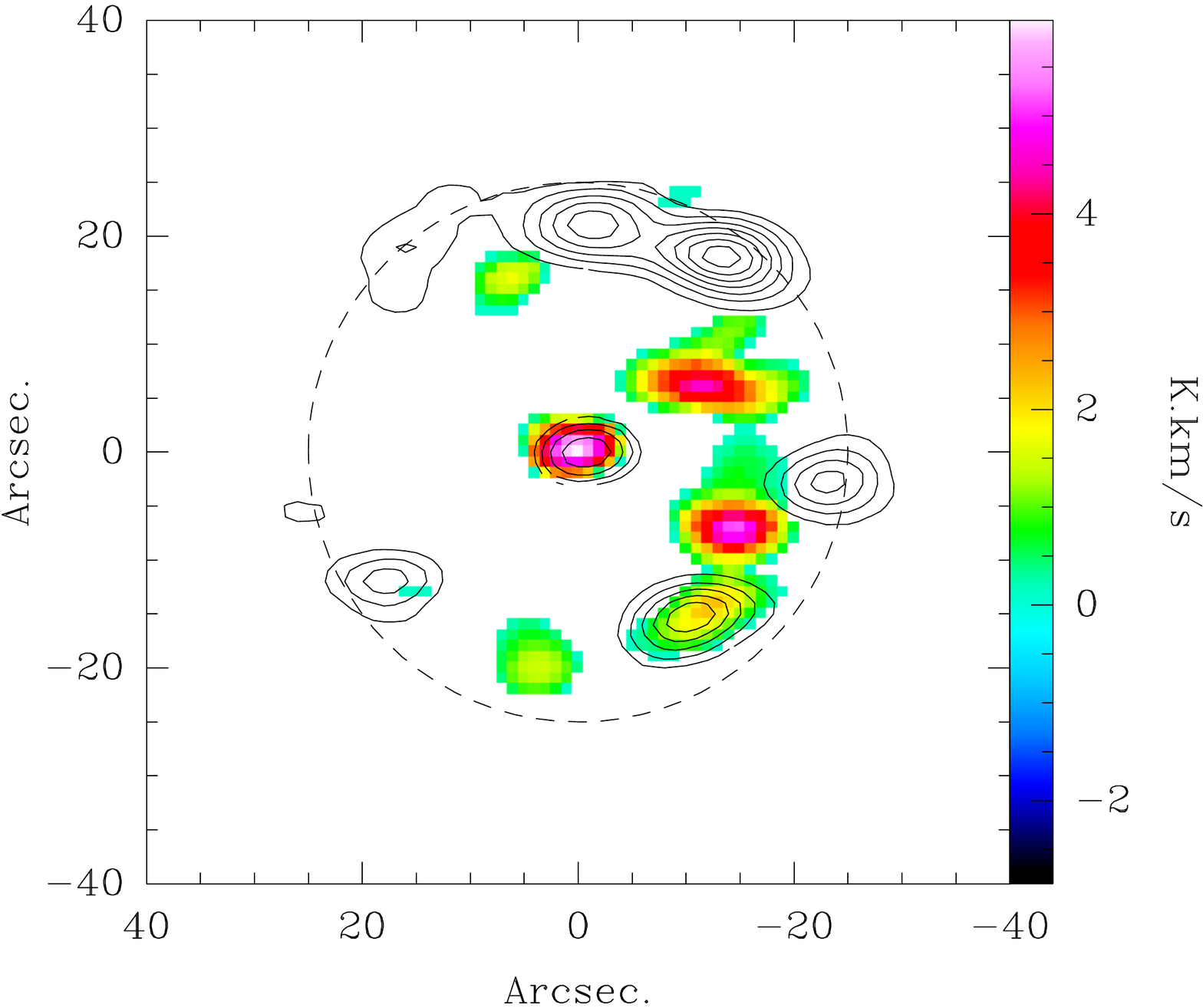} 
  \end{center}
 \caption[]{In color: Integrated CO 2-1 map between +90 and +110 \kms, showing the western arc. The contours are taken from Fig.~\ref{fig:mom0}{\it d} and show the clumpy CO ring, and the dust shell seen with {\it Herschel} is represented by a dashed circle with 25$''$ radius. As can be seen, the brightest arc emission is not coincident with the CO ring. 
\label{fig:arc}}
\end{figure}

Between +90 and +110 \kms\ there is a western arc feature, which is not spatially connected with the spur emission. It is also not coincident with the clumpy CO ring (see Fig.~\ref{fig:arc}). This feature is redshifted by $\sim$70-90 \kms\ with respect to the systemic velocity of the star, and has no blueshifted counterpart. Its velocity is also far removed from Galactic rotation at any distance towards \Egg, suggesting it is not interstellar but instead connected with the star. However, the origin and nature of this western arc is unknown. 

The western arc has an integrated intensity of $\sim$8~K\,\kms\ between 90 and 110 \kms. Using the method from Sect.~\ref{sect:ring} to roughly estimate the mass of this feature, we find $N_{CO}$=10$^{16}$~cm$^{-2}$, and, assuming spherical clumps of size 8$''$, estimate a total mass of $\sim$0.09~M$_\odot$ in three clumps.

\section{Discussion}

The ACA data reveal for the first time the distribution of the molecular gas in the Fried Egg nebula and allow us to reconcile the previous APEX spectra with optical and {\it Herschel} data.
They also reveal new features, like the kinematics of the clumpy CO ring and the strong asymmetry of the spur structure, that call for further investigation. 

\subsection{Origin of the clumpy CO ring}

The clumpy CO ring is formed by a collection of apparently individual clumps which collectively have {\em i)}  a coherent kinematic structure, and {\em ii)} an roughly circular symmetry with respect to the star, which {\em iii)} spatially matches the $\sim$60~K dust emission seen with {\it Herschel}. The symmetry strongly implies an origin connected to the star and, as discussed in Sect.~\ref{sect:ring}, the kinematics are not compatible with an expanding shell. From the velocity structure the ring instead resembles a torus, close to the plane of the sky (inclination $\sim$35$^\circ$, assuming circular symmetry).
From the velocities of the clumps we estimate the ejection velocity is between 20 and $\sim$50 \kms, meaning the torus was ejected between $\sim$7000 and 20\,000 years ago, during the previous RSG phase of the star. We note that the lifetime of unshielded CO against the average interstellar UV-field is only $\sim$100 years \citep{van-Dishoeck:1988aa}. This implies that we are seeing only the densest inhomogeneities in the torus, where the CO is sufficiently shielded, which may explain the clumpy appearance. 

This clumpy torus resembles the famous ring around the supernova SN~1987A, which is believed to have been ejected during the progenitor star's RSG phase \citep{Smith:2007aa}. Several Galactic supergiants have also been found to have similar morphologies, with an equatorial ring and a bipolar hourglass nebula, e.g., SBW1 \citep{Smith:2013aa} and Sher 25 \citep{Brandner:1997aa}. While \Egg\ does not appear to have an accompanying bipolar nebula, it is possible that it is no longer detectable in CO due to dissociation. There has also been some suggestion of an embedded disk or torus in the YHG IRC+10420 \citep{Castro-Carrizo:2007aa}.
The main proposed formation scenarios for the torus are either a fast, low-density wind expanding into a previous slow, dense wind with a strong equatorial density enhancement, or a burst of intense mass-loss from a rapidly rotating star. 
For a relatively large and cool star like \Egg, acquiring a sufficiently large rotation rate would probably require interaction or merger with a binary companion. 

It is possible that the clumpy CO ring was not ejected by the star, but is instead swept up ISM material. The isotropic wind from the star could create a cavity in the ISM, pushing ISM clouds into a ring shape. However, the lack of an expanding spherical geometry would require the ISM to be significantly denser in a plane around the star before being swept up. 
Furthermore, the kinematic structure of the clumpy ring must be considered. The clumps show redshifted velocities to the south-east and blue-shifted velocities to the north-west, and there are clumps at the systemic velocity to the south-west and north-east. This might be explained by the star moving towards the south-east, pushing ISM ahead of it, though this is difficult to reconcile with the circular symmetry of the ring. 
Another aspect is the inter-clump velocity difference of $\sim$20 \kms. It is possible that the swept-up ISM clouds could retain their relative velocities, resulting in the observed clumps at various velocities at the same distance from the star. However, typical ISM intercloud velocities are $\sim$5$-$10~\kms\ (see e.g. \citealp{Malhotra:1994aa}, and references therein), a factor of 2-4 lower than the observed value.

From the current data it is likely that the clumpy CO ring is a torus ejected by the star, the formation of which may suggest the existence of a binary companion. It is also possible that the clumpy CO ring is swept-up ISM material, but the maintenance of symmetry and a coherent kinematical structure between the various clumps is difficult to explain in this case. The kinematics also do not rule out the possibility of infalling material, but this is unlikely given the large scales involved: the CO ring is about 40$''$$\approx$0.8~pc across.
Further observations, for example allowing for the determination of the $^{12}$C/$^{13}$C ratio or the properties of the dust, might be able to elucidate the origin of this ring.

\subsection{Origin of the spur}

The spur is an unexpected feature, but we argue that it is associated with the star, for several reasons. First, it appears to spatially extend from the star to the clumpy CO ring, which would be an unlikely coincidence for an ISM cloud in the line of sight. Second, the velocity of the spur is offset by at least $\sim$ 60 \kms\ from Galactic rotation at any distance between 0 and 10 kpc towards IRAS~17163. Third, a corresponding feature can be seen in the {\it Herschel} dust image from \citet{Hutsemekers:2013aa}, as shown in Fig.~\ref{fig:contours}, in the form of an elongation in the direction of the CO spur. This suggests that the spur and clumpy CO ring have similar properties, including a dust temperature of $\sim$60~K. Furthermore, the estimated mass of the spur and clumpy CO ring, $\sim$3--5~\Msun, is roughly similar to the total gas mass estimated by \citet{Hutsemekers:2013aa} of $\sim$7~\Msun. And finally, the $^{12}$CO/$^{13}$CO ratio of $13^{+10}_{-5}$ is consistent with material processed by a RSG star. There is also some evidence of a unidirectional ejection in the YHG AFGL~2343 \citep{Quintana-Lacaci:2008aa}.

To create a feature like the spur requires a break in symmetry. 
One potential explanation is that the star has undergone sustained, directed mass-loss on a timescale of a few 10$^3$ years (assuming an expansion velocity of 100~\kms), although the mechanism is uncertain. 
\citet{OGorman:2015aa} found a $2.5\times10^{-4}$~M$_\odot$ dust feature offset by 400~AU from the red supergiant VY~CMa, and suggested a localised, long-lived magnetohydrodynamic disturbance could cause such directed mass-loss from the star. However, the formation timescale for this dust feature is only $\sim$20~years, and such localised mass-loss would appear as random mass ejections on larger spatial scales. The spur of IRAS~17163 has an extent of $>$8$\times 10^{4}$~AU, corresponding to a timescale of $>$3700 years (assuming a velocity of 100 \kms), so it is difficult to explain in the same way. The spur is also much more massive: $\sim$3~\Msun\ (Sect.~\ref{sect:spur}).

Is it possible that the spur is an interstellar cloud, with a large peculiar velocity and aligned with the star by chance? In this case, we can roughly estimate its linear size and hence its distance using Larson's Law \citep{Larson:1981aa} of turbulence in molecular clouds: $\sigma_v = 1.1 \times l^{0.38}$ where $\sigma_v$ is the velocity dispersion (linewidth) in \kms\ and $l$ the cloud size in pc. Using a linewidth of 8 \kms\ this gives a huge linear size of $\sim$185~pc, and combined with the spur's angular size of 15$''$ this implies a distance of 2500~kpc. Hence, the spur is very unlikely to be a molecular cloud.

\section{Conclusions}

Our knowledge of the mass-loss, evolution and surroundings of post-RSG stars is very limited.
We report here observations of the circumstellar emission around the yellow hypergiant star
IRAS 17163 with the ALMA Compact Array (ACA). In CO 2-1, a spectrum taken at the star shows the broad faint plateau of a stellar wind, centered on +18~\kms. Hence we resolve the apparent velocity discrepancy previously found from CO and optical emission lines \citep{Wallstrom:2015aa}. The stellar wind has a large expansion velocity of 100$\pm$10 \kms\, and a mass-loss rate of 8$\pm$1.5 $\times10^{-5}$~M$_\odot$\,yr$^{-1}$, under the assumption of a constant and isotropic wind. We also detect the H30$\alpha$ recombination line of hydrogen towards the star. The line is centered on 21$\pm$3~\kms\ with a FWHM of 57$\pm$6~\kms.

Our ACA CO observations reveal a complex circumstellar environment, contrasting with the symmetry of previous infrared images of the star. We detect the CO counterpart to the previously observed {\it Herschel} dust shell, in the form of a clumpy CO ring at $v_{sys}\pm$20 \kms. However, this ring does not have the velocity structure of an expanding shell, rather it shows a strong velocity gradient with a position angle $\sim$100$^\circ$. The structure resembles a torus, like the one famously seen around the supernova SN 1987A. The probable ejection scenarios for this torus suggests \Egg\ may have a binary companion.
Further breaking the symmetry is an elongated spur, stretching between the central unresolved stellar wind and the CO ring, which peaks at a velocity around +40 \kms\ with respect to the star. There is no obvious blueshifted counterpart to this feature on the other side of the star. The $^{12}$CO/$^{13}$CO ratio in the spur is measured to be $13^{+10}_{-5}$, suggesting it was ejected by the star in its previous RSG phase. 

These observations demonstrate the complexity of massive evolved stars. ALMA is now allowing us to study the morphology of their circumstellar envelopes, the gas kinematics, and their mass-loss prior to a supernova explosion.

\begin{acknowledgements}
We would like to thank the referee for their constructive comments which helped improve the quality and clarity of this manuscript. \\
This paper makes use of the following ALMA data:
    ADS/JAO.ALMA\#2013.1.00502.S. ALMA is a partnership of ESO (representing
    its member states), NSF (USA) and NINS (Japan), together with NRC
    (Canada) and NSC and ASIAA (Taiwan) and KASI (Republic of Korea), in
    cooperation with the Republic of Chile. The Joint ALMA Observatory is
    operated by ESO, AUI/NRAO and NAOJ. \\
NC and GQL acknowledge that their research leading to these results has
received funding from the European Research Council under the European
Union's Seventh Framework Programme (FP/2007-2013) / ERC Grant Agreement
n. 610256 (NANOCOSMOS). \\
EL and RSz were partially supported by the National Research Center, Poland, DEC-2013/08/M/ST9/00664, within the framework of the HECOLS International Associated Laboratory. \\
AZ was supported by the Science and Technology Research Council through grant ST/L000768/1. \\
HO acknowledges financial support from the Swedish Research Council.
\end{acknowledgements}

\bibliography{papers}

\begin{thebibliography}{27}
\expandafter\ifx\csname natexlab\endcsname\relax\def\natexlab#1{#1}\fi

\bibitem[{{Arnal} {et~al.}(2008){Arnal}, {Duronea}, \&
  {Testori}}]{Arnal:2008aa}
{Arnal}, E.~M., {Duronea}, N.~U., \& {Testori}, J.~C. 2008, \aap, 486, 807

\bibitem[{{Bernes}(1979)}]{Bernes:1979aa}
{Bernes}, C. 1979, \aap, 73, 67

\bibitem[{{Brandner} {et~al.}(1997){Brandner}, {Grebel}, {Chu}, \&
  {Weis}}]{Brandner:1997aa}
{Brandner}, W., {Grebel}, E.~K., {Chu}, Y.-H., \& {Weis}, K. 1997, \apjl, 475,
  L45

\bibitem[{{Castro-Carrizo} {et~al.}(2007){Castro-Carrizo}, {Quintana-Lacaci},
  {Bujarrabal}, {Neri}, \& {Alcolea}}]{Castro-Carrizo:2007aa}
{Castro-Carrizo}, A., {Quintana-Lacaci}, G., {Bujarrabal}, V., {Neri}, R., \&
  {Alcolea}, J. 2007, \aap, 465, 457

\bibitem[{{de Jager}(1998)}]{de-Jager:1998aa}
{de Jager}, C. 1998, \aapr, 8, 145

\bibitem[{{Dinh-V.-Trung} {et~al.}(2009){Dinh-V.-Trung}, {Muller}, {Lim},
  {Kwok}, \& {Muthu}}]{Dinh-V.-Trung:2009aa}
{Dinh-V.-Trung}, {Muller}, S., {Lim}, J., {Kwok}, S., \& {Muthu}, C. 2009,
  \apj, 697, 409

\bibitem[{{Gonzalez} \& {Wallerstein}(2000)}]{Gonzalez:2000aa}
{Gonzalez}, G. \& {Wallerstein}, G. 2000, \aj, 119, 1839

\bibitem[{{Hutsem{\'e}kers} {et~al.}(2013){Hutsem{\'e}kers}, {Cox}, \&
  {Vamvatira-Nakou}}]{Hutsemekers:2013aa}
{Hutsem{\'e}kers}, D., {Cox}, N.~L.~J., \& {Vamvatira-Nakou}, C. 2013, \aap,
  552, L6

\bibitem[{{Lagadec} {et~al.}(2011){Lagadec}, {Zijlstra}, {Oudmaijer},
  {Verhoelst}, {Cox}, {Szczerba}, {M{\'e}karnia}, \& {van
  Winckel}}]{Lagadec:2011aa}
{Lagadec}, E., {Zijlstra}, A.~A., {Oudmaijer}, R.~D., {et~al.} 2011, \aap, 534,
  L10

\bibitem[{{Larson}(1981)}]{Larson:1981aa}
{Larson}, R.~B. 1981, \mnras, 194, 809

\bibitem[{{Malhotra}(1994)}]{Malhotra:1994aa}
{Malhotra}, S. 1994, \apj, 433, 687

\bibitem[{{O'Gorman} {et~al.}(2015){O'Gorman}, {Vlemmings}, {Richards},
  {Baudry}, {De Beck}, {Decin}, {Harper}, {Humphreys}, {Kervella}, {Khouri}, \&
  {Muller}}]{OGorman:2015aa}
{O'Gorman}, E., {Vlemmings}, W., {Richards}, A.~M.~S., {et~al.} 2015, \aap,
  573, L1

\bibitem[{{Origlia} {et~al.}(2013){Origlia}, {Oliva}, {Maiolino},
  {Mucciarelli}, {Baffa}, {Biliotti}, {Bruno}, {Falcini}, {Gavriousev},
  {Ghinassi}, {Giani}, {Gonzalez}, {Leone}, {Lodi}, {Massi}, {Montegriffo},
  {Mochi}, {Pedani}, {Rossetti}, {Scuderi}, {Sozzi}, \&
  {Tozzi}}]{Origlia:2013aa}
{Origlia}, L., {Oliva}, E., {Maiolino}, R., {et~al.} 2013, \aap, 560, A46

\bibitem[{{Oudmaijer} {et~al.}(2009){Oudmaijer}, {Davies}, {de Wit}, \&
  {Patel}}]{Oudmaijer:2009aa}
{Oudmaijer}, R.~D., {Davies}, B., {de Wit}, W.-J., \& {Patel}, M. 2009, in
  Astronomical Society of the Pacific Conference Series, Vol. 412, The Biggest,
  Baddest, Coolest Stars, ed. D.~G. {Luttermoser}, B.~J. {Smith}, \& R.~E.
  {Stencel}, 17

\bibitem[{{Quintana-Lacaci} {et~al.}(2008){Quintana-Lacaci}, {Bujarrabal}, \&
  {Castro-Carrizo}}]{Quintana-Lacaci:2008aa}
{Quintana-Lacaci}, G., {Bujarrabal}, V., \& {Castro-Carrizo}, A. 2008, \aap,
  488, 203

\bibitem[{{Reid} {et~al.}(2009){Reid}, {Menten}, {Zheng}, {Brunthaler},
  {Moscadelli}, {Xu}, {Zhang}, {Sato}, {Honma}, {Hirota}, {Hachisuka}, {Choi},
  {Moellenbrock}, \& {Bartkiewicz}}]{Reid:2009aa}
{Reid}, M.~J., {Menten}, K.~M., {Zheng}, X.~W., {et~al.} 2009, \apj, 700, 137

\bibitem[{{Sch{\"o}ier} \& {Olofsson}(2001)}]{Schoier:2001ab}
{Sch{\"o}ier}, F.~L. \& {Olofsson}, H. 2001, \aap, 368, 969

\bibitem[{{Smith} {et~al.}(2013){Smith}, {Arnett}, {Bally}, {Ginsburg}, \&
  {Filippenko}}]{Smith:2013aa}
{Smith}, N., {Arnett}, W.~D., {Bally}, J., {Ginsburg}, A., \& {Filippenko},
  A.~V. 2013, \mnras, 429, 1324

\bibitem[{{Smith} {et~al.}(2007){Smith}, {Bally}, \&
  {Walawender}}]{Smith:2007aa}
{Smith}, N., {Bally}, J., \& {Walawender}, J. 2007, \aj, 134, 846

\bibitem[{{Szczerba} {et~al.}(2007){Szczerba}, {Si{\'o}dmiak}, {Stasi{\'n}ska},
  \& {Borkowski}}]{Szczerba:2007aa}
{Szczerba}, R., {Si{\'o}dmiak}, N., {Stasi{\'n}ska}, G., \& {Borkowski}, J.
  2007, \aap, 469, 799

\bibitem[{{van der Tak} {et~al.}(2007){van der Tak}, {Black}, {Sch{\"o}ier},
  {Jansen}, \& {van Dishoeck}}]{van-der-Tak:2007aa}
{van der Tak}, F.~F.~S., {Black}, J.~H., {Sch{\"o}ier}, F.~L., {Jansen}, D.~J.,
  \& {van Dishoeck}, E.~F. 2007, \aap, 468, 627

\bibitem[{{van Dishoeck} \& {Black}(1988)}]{van-Dishoeck:1988aa}
{van Dishoeck}, E.~F. \& {Black}, J.~H. 1988, \apj, 334, 771

\bibitem[{{Vassilev} {et~al.}(2008){Vassilev}, {Meledin}, {Lapkin}, {Belitsky},
  {Nystr{\"o}m}, {Henke}, {Pavolotsky}, {Monje}, {Risacher}, {Olberg},
  {Strandberg}, {Sundin}, {Fredrixon}, {Ferm}, {Desmaris}, {Dochev},
  {Pantaleev}, {Bergman}, \& {Olofsson}}]{Vassilev:2008aa}
{Vassilev}, V., {Meledin}, D., {Lapkin}, I., {et~al.} 2008, \aap, 490, 1157

\bibitem[{{Wallstr{\"o}m} {et~al.}(2015){Wallstr{\"o}m}, {Muller}, {Lagadec},
  {Black}, {Oudmaijer}, {Justtanont}, {van Winckel}, \&
  {Zijlstra}}]{Wallstrom:2015aa}
{Wallstr{\"o}m}, S.~H.~J., {Muller}, S., {Lagadec}, E., {et~al.} 2015, \aap,
  574, A139

\bibitem[{{Wilson} \& {Rood}(1994)}]{Wilson:1994aa}
{Wilson}, T.~L. \& {Rood}, R. 1994, \araa, 32, 191

\bibitem[{{Yang} {et~al.}(2010){Yang}, {Stancil}, {Balakrishnan}, \&
  {Forrey}}]{Yang:2010aa}
{Yang}, B., {Stancil}, P.~C., {Balakrishnan}, N., \& {Forrey}, R.~C. 2010,
  \apj, 718, 1062

\bibitem[{{Zijlstra} {et~al.}(2001){Zijlstra}, {Chapman}, {te Lintel Hekkert},
  {Likkel}, {Comeron}, {Norris}, {Molster}, \& {Cohen}}]{Zijlstra:2001aa}
{Zijlstra}, A.~A., {Chapman}, J.~M., {te Lintel Hekkert}, P., {et~al.} 2001,
  \mnras, 322, 280

\end{thebibliography}
\bibliographystyle{aa.bst}





\newpage
\onecolumn 
\appendix

\section{CO 2-1 channel map}

\begin{figure}[h!]
 \caption[]{ALMA ACA channel map of the CO 2-1 around \Egg. The velocity of each channel is given in the top left; note the systemic velocity of the star is +18 \kms. The cool dust ring seen with {\it Herschel} is represented by a dashed circle with 25$''$ radius, and the location of the star, at R.A. 17:19:49.33 Dec. $-$39:10:37.94, is marked with a small cross. The restoring beam is shown as a white ellipse in the bottom left of the bottom left image. The common color scale is shown on the bottom right image. The figure continues on the next page.
\label{fig:chanmap}}
  \begin{center}
    \includegraphics[width=\textwidth]{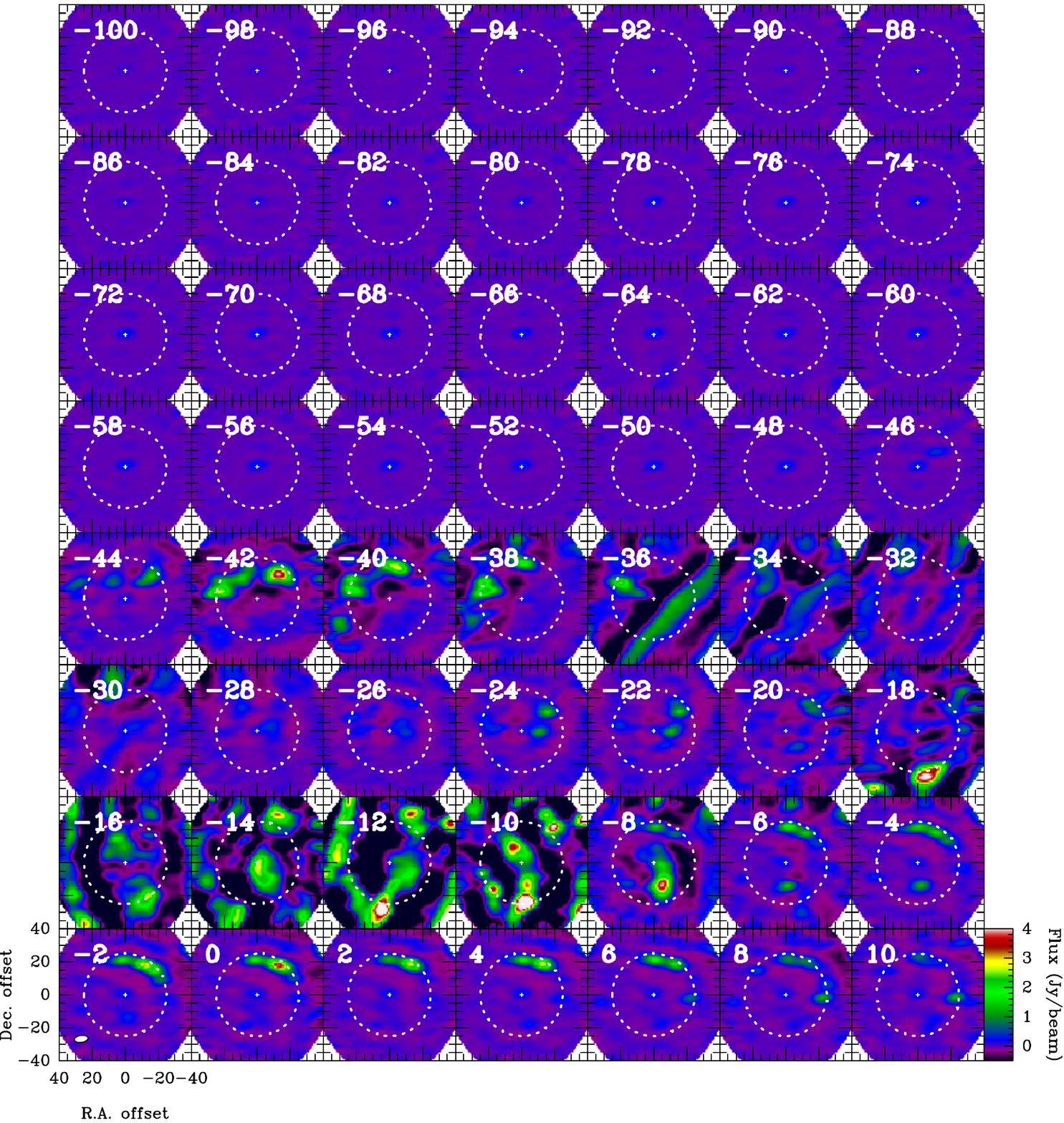}
  \end{center}
\end{figure}

\begin{figure*}[h!]
  \begin{center}
    \includegraphics[width=\textwidth]{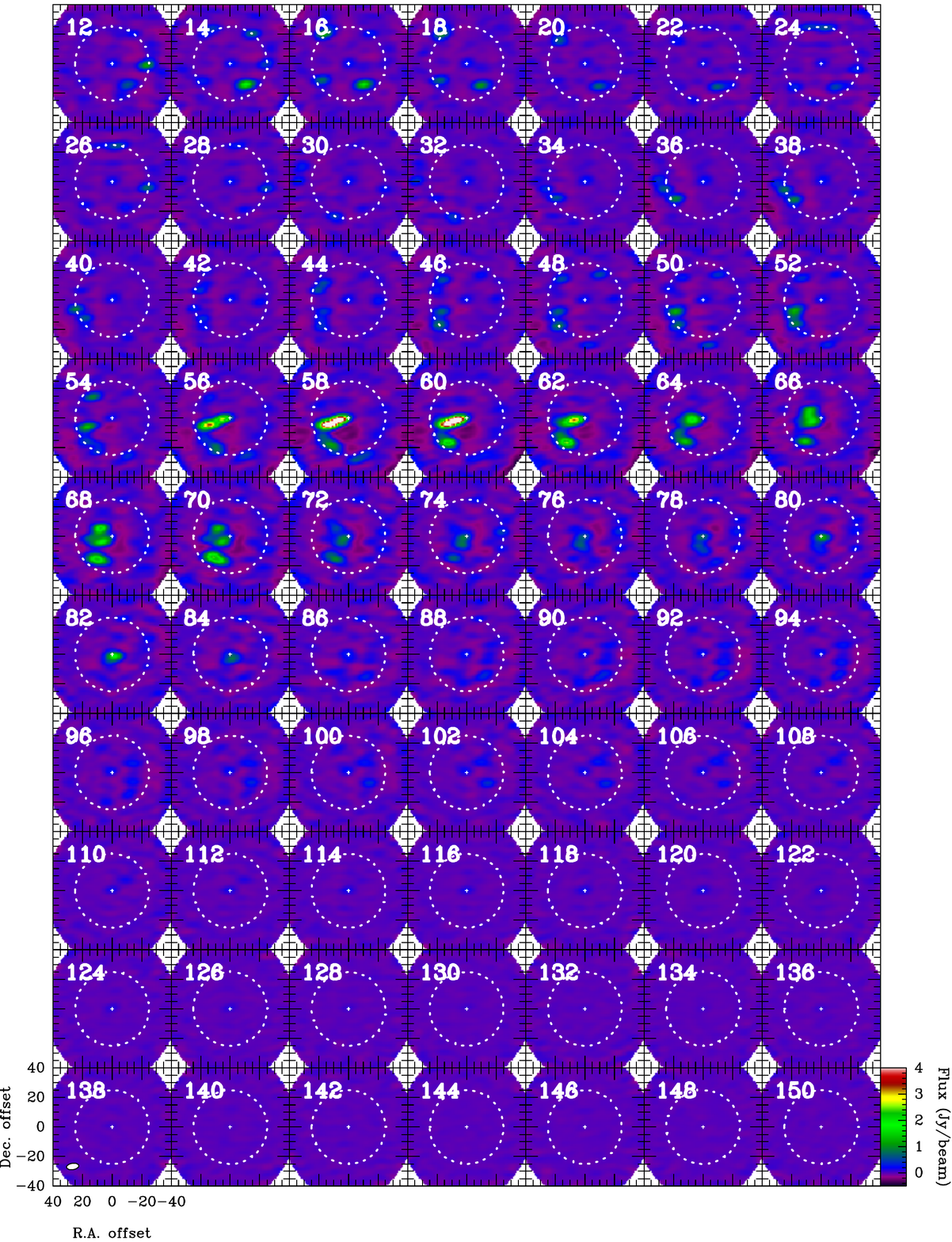}
  \end{center}
\end{figure*}

\end{document}